\documentclass[letterpaper, 10 pt, conference]{ieeeconf}  

\IEEEoverridecommandlockouts                              

\usepackage{adjustbox}
\usepackage{amsfonts}
\usepackage{amsmath}
\usepackage{booktabs}
\usepackage{graphicx}
\usepackage{import}
\usepackage{multirow}
\usepackage{subcaption}
\usepackage{tabularx} 
\usepackage{transparent}
\usepackage{url}
\usepackage{xcolor}

\DeclareMathOperator*{\argmin}{arg\,min}

\newcommand{\norm}[1]{\left\lVert#1\right\rVert}

\graphicspath{{images/}}

\title{\LARGE \bf
Guided Depth Upsampling for Precise Mapping of Urban Environments
}

\author{Sascha Wirges$^{1}$, Bj\"orn Roxin$^{2}$ , Eike Rehder$^{2}$, Tilman K\"uhner$^{1}$ and Martin Lauer$^{2}$
\thanks{$^{1}$Sascha Wirges and Tilman K\"uhner are with the FZI Research Center for Information Technology and Karlsruhe Institute of Technology, Germany.
        {\tt\small \{wirges,kuehner\}@fzi.de}}%
\thanks{$^{2}$Bj\"orn Roxin, Eike Rehder and Martin Lauer are with the Institute of Measurement and Control, Karlsruhe Insitute of Technology, Germany
        {\tt\small roxinbj@gmail.com,\{rehder,lauer\}@kit.edu}}%
}

\begin{document}

\maketitle
\thispagestyle{empty}
\pagestyle{empty}

\begin{abstract}

We present an improved model for MRF-based depth upsampling, guided by image- as well as 3D surface normal features.
By exploiting the underlying camera model we define a novel regularization term that implicitly evaluates the planarity of arbitrary oriented surfaces.
Our method improves upsampling quality in scenes composed of predominantly planar surfaces, such as urban areas.
We use a synthetic dataset to demonstrate that our approach outperforms recent methods that implement distance-based regularization terms.
Finally, we validate our approach for mapping applications on our experimental vehicle.

\end{abstract}


\section{Introduction}
\label{sec:introduction}

Perception and localization algorithms developed for automated driving tasks rely on accurate environment models.
These models are usually generated using information provided by mobile sensors such as cameras or range sensors.
Whereas cameras provide 2D projections of surface reflectances with high spatial resolution, range sensors usually provide precise 3D surface positions.
However, the spatial resolution of modern range sensors is sparse compared to cameras.

Currently, most systems perform environmental mapping within one sensor domain which has several drawbacks.
Common methods usually perform feature estimation and matching to find corresponding surface landmarks between subsequent measurement frames.
For camera-based mapping methods, the scale might be either subject to drift or hard to estimate accurately in the calibration process which results in globally inconsistent maps.
For range sensor-based methods the resulting map may consist of accurate but spatially sparse 3D points which inherently induces errors on surface feature estimation and reconstruction.
Thus, our aim is to combine the strengths of both sensor types to generate a map that consists of spatially dense surface features.

Here, we propose a guided depth upsampling method that estimates surfaces accurately for each camera pixel within scenes composed of predominantly planar surfaces, such as urban areas.
Provided with a calibrated camera-laser setup, the 3D surface point position can be determined by evaluating the viewing ray corresponding to an image coordinate at an estimated depth.

However, as different image areas usually have varying 3D point densities, the quality of depth upsampling might vary drastically.
Therefore, we are also interested in finding a confidence measure for each depth estimate.
We show that our method is capable of performing accurate upsampling within image areas that contain only few 3D point observations.
Finally, we provide a filtering method that stems from our optimization model to filter out ill-conditioned depths.

\begin{figure}[t]
\centering
\includegraphics[width=\linewidth]{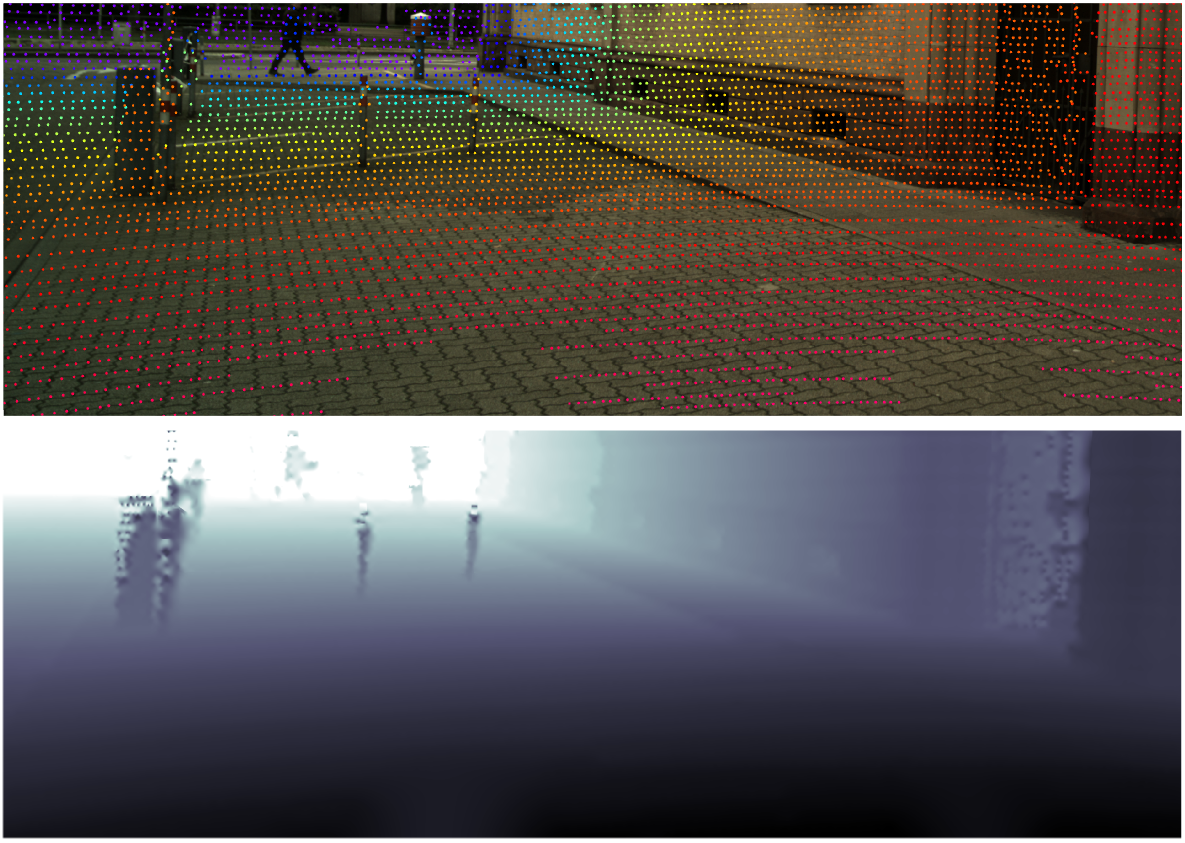}
\caption{Top: Input RGB image with range sensor data input overlay.
Bottom: Upsampled high-resolution depth image}
\label{fig:upsampled_real_scene} 
\end{figure}

By describing similarities and differences of related work in guided depth upsampling in section \ref{sec:motivation_and_related_work}, we show common drawbacks and emphasize our ideas to overcome these problems.
Based on these findings, we formalize our objectives in depth upsampling and derive the underlying Markov Random Field model in section \ref{sec:guided_depth_upsampling}.
We will then validate our approach on a photorealistic indoor dataset and our experimental vehicle (section \ref{sec:applications_and_experiments}).
Finally, we conclude our findings in section \ref{sec:conclusion} and show our next plans in guided depth upsampling.

\section{Motivation and Related Work}
\label{sec:motivation_and_related_work}
Our general objective in depth upsampling is to estimate depths $\hat{d}_i$ for each image coordinate $i \in \mathcal{I}$ of the image $\mathcal{I}$.
Depth observations $d_{j, \text{obs}}, j \in \mathcal{O} \subset \mathcal{I}$ from range sensors might be available only for a small subset $\mathcal{O}$ of image coordinates.

Assuming a calibrated camera-laser setup, each $\hat{d}_i$ can be transformed into a corresponding 3D point
\begin{equation} \label{eq:point_from_depth}
\hat{\mathbf{p}}_{i} = \mathbf{ray}_i(\hat{d}_i) = \mathbf{ori}_{i} + \hat{d}_i \cdot \mathbf{dir}_{i}
\end{equation}
using the viewing ray function $\hat{\mathbf{ray}}_i$ that includes for each image coordinate $i$ the direction $\mathbf{dir}_{i}$ and the viewpoint $\mathbf{ori}_{i}$ of the respective line of sight.
In the following, we will use $\hat{\mathbf{p}}_{i}$ and $\hat{d}_i$ interchangeably.

Upsampling methods may be divided into local or global methods.
Whereas local methods \cite{Kopf2007}, \cite{Liu2013} can be used for upsampling images with mostly dense and uniformly distributed depth observations, global methods show better performance on data with sparse and non-uniformly distributed observations.
In urban mapping, however, the number of observations might vary drastically, depending on the scene setting.
Therefore, we focus on global upsampling methods.

Within global methods, the optimal depths $d^*_i, i \in \mathcal{I}$ arranged in
\begin{equation} \label{eq:optim_problem}
\mathbf{d}^* = \argmin_{\hat{\mathbf{d}}} \Phi(\hat{\mathbf{d}})
\end{equation}
minimize the cost function $\Phi$ that may be composed of various cost terms.
\cite{Thrun2005} models this problem as a Markov Random Field with a cost term
\begin{equation} \label{eq:mrf_thrun}
\Phi = \Phi_{\text{data}} + \Phi_{\text{reg}},
\end{equation}
that does not only minimize costs towards the given observation data, but also within the direct neighborhood $\mathcal{N}_i$ of each image coordinate $i$, where the regularization term
\begin{equation} \label{eq:regularization_thrun}
\Phi_{\text{reg}} = \sum_{i \in \mathcal{I}} \sum_{n \in \mathcal{N}_i} w_{i n} (\hat{d}_i - \hat{d}_n)^2
\end{equation}
is used to enforce that estimated depths of direct image coordinate neighbors (e.g. within a 4-connected grid) are similar.
However, their model assumption does not hold for arbitrary planes as it regularizes towards similar depths.

The weights $w_{i n}$ in equation (\ref{eq:regularization_thrun}) might be used to include additional information on the problem.
Whereas invariant weights are used in image filtering applications \cite{chen2014fast}, weights depending on image features can guide upsampling and thus improve quality.
Moreover, in all guided approaches, image features are used to indicate depth discontinuities (see table \ref{tab:related_work}).
In particular, \cite{lee01} shows that image values and range measurements share second order statistics.
Based on this work, either gray scale \cite{Thrun2005} or color intensity gradients are used.
\cite{Schneider2016} includes semantic information in the regularization term and determines extended neighborhoods based on geodesic distances.
Even higher-order terms such as the anisotropic diffusion tensor \cite{Ferstl} or \cite{Park2011} might be used.
The authors add a non-local means regularization term, which uses an anisotropic structural-aware filter to allow similar pixels in extended neighborhoods to reinforce each other.

Although guided approaches based on image features have been studied extensively,
a major drawback of existing methods is the lack of incorporating 3D features into the upsampling process.
Therefore, we show the benefit of including 3D surface normals into our problem.
\begin{table}
\begin{adjustbox}{width=\linewidth}
\def\arraystretch{1.2}
\begin{tabular}[3pt]{ | c  | c | c | c | c | }
\hline
  									&  \cite{Thrun2005} 	& \cite{Park2011} 	& \cite{Ferstl}	& \cite{Schneider2016}	\\ \hline
    RGB / gray scale values  		& x 					& x					& x 				& x 						\\ \hline 
    Spatial distance 				&   					& x 					&   				& x 						\\ \hline
    Anisotropic diffusion tensor 	& 					& x 					& x 				&   						\\ \hline
    Semantic information 			& 					&  					& 				& x 						\\ \hline
\end{tabular}
\end{adjustbox}
\caption{Image features used in different contributions} \label{tab:related_work}
\end{table}

Even if recent methods achieve accurate results, they do not account for confidences in the estimation problem.
We provide a simple method based on estimating the parameter covariance of the underlying optimization problem at the end of the next section.

\section{Guided Depth Upsampling}
\label{sec:guided_depth_upsampling}

For each image coordinate $i$, we aim to determine its depth $\hat{d}_i$ and a depth confidence measure $\sigma_i$.
To achieve this, we require a calibrated camera-laser rig that provides viewing ray lookup functions $\mathbf{ray}_i$ as described in equation (\ref{eq:point_from_depth}) and the transform $\mathbf{p}_{ext} \in SE(3)$ between range sensor and camera frame to be known.
Given $\mathbf{p}_{ext}$, observed 3D point features $\mathbf{f}_j, j \in \mathcal{O}$ can be transformed into the camera frame and mapped to the image coordinate $j$.

As in equation (\ref{eq:mrf_thrun}) we model our upsampling problem as a Markov Random Field containing data costs $\Phi_{\text{data}}$ and regularization costs $\Phi_{\text{reg}}$.
We can include additional image features into the optimization problem which we explain in section \ref{subsec:weights}.
These cost terms should be minimized starting from depth priors $\hat{d}_{i, 0}$ determined by our initialization strategy explained in section \ref{subsec:prior_estimation}.
In the following, we describe the different energy functions included in our model.

\subsection{Data Costs}

For each observation $d_{j, \text{obs}}, j \in \mathcal{O}$ we set up data costs
\begin{equation}\label{eq:data_cost}
\Phi_{\text{data}} = \sum_{j \in \mathcal{O}} w_{\text{data}} (\phi_{j, \text{depth}}^2 + \sum_{n \in \mathcal{N}_j } \phi_{j n, \text{normal}}^2)
\end{equation}
weighted by $w_{\text{data}}$.
Here, $\mathcal{N}_j$ is the direct neighborhood of image coordinate $j$ which we choose to be a 4-connected grid.

Since we want to include depth observations from range sensors, depth residuals
\begin{equation}\label{eq:depth_residual}
\phi_{j, \text{depth}} = \hat{d}_j - d_{j, \text{obs}}
\end{equation}
evaluate the difference between estimated and observed depths for each image coordinate $j \in \mathcal{O}$.

In addition, we include estimated surface normals from range sensors as pseudo measurements into our problem.
Therefore, normal residuals
\begin{equation}\label{eq:normal_residual}
\phi_{j n, \text{normal}} = \mathbf{n}_{P_j}^T \hat{\mathbf{p}}_n - d_{P_j}
\end{equation}
evaluate the signed point-to-plane distance between the constructed plane $P_j$ and the point $\hat{\mathbf{p}}_n = \mathbf{ray}_n(\hat{d}_n)$.
The plane is constructed from the surface normal $\mathbf{n}_{\text{obs}}$ and can be expressed in normal form
\begin{equation}\label{eq:hnf}
\mathbf{n}_{P_j}^T, \mathbf{x} - d_{P_j} = 0.
\end{equation}

\subsection{Regularization Costs}

To model coupling in Markov sense, we add regularization cost terms for each image coordinate $i$ within its direct neighborhood $\mathcal{N}_{i}$.
Whereas 8-connected grids provide a better coupling with a large number of residual blocks which decreases optimization speed, choosing two neighbors will lead to poor coupling and decrease convergence.
Thus, we choose 4-connected grids as they provide a good trade-off between coupling and the amount of coupling residuals in the problem.

We aim to minimize the regularization cost
\begin{equation}\label{eq:regularization_cost}
\Phi_{\text{reg}} = \sum_{i \in \mathcal{I}}\sum_{\mathcal{D} \subset \mathcal{N}_i } w_i(\mathcal{D}) \phi_{i, \text{planar}}(\mathcal{D})^2
\end{equation}
for each image coordinate $i$ in the image $\mathcal{I}$.
Here, $w_i(\mathcal{D})$ is a weighting term depending on $i$ and a subset $\mathcal{D}$ of the neighborhood $\mathcal{N}_i$.
We will explain in section \ref{subsec:weights} how $w_i(\mathcal{D})$ is composed.
In the simplest case the residual terms $\phi_{i, \text{planar}}$ are evaluated between pairs of image coordinates $(i, n)$ as presented in equation (\ref{eq:regularization_thrun}), where $n \in \mathcal{N}_i$.
Here, we extend the residual computation to be dependent on sets $\mathcal{D} \subset \mathcal{N}_i$ of multiple image coordinates.

Instead of regularizing towards constant depth (e.g. as in \cite{Thrun2005}), we enforce the surface points to be coplanar.
Thus, we aim to find an appropriate residual term that shows good convergence properties.

One option would be to estimate surface normals explicitly based on all points in the corresponding neighborhood $\mathcal{N}_i$ and find an appropriate point-to-plane residual, similar to equation (\ref{eq:normal_residual}).

\begin{figure}[htbp]
\centering \def\svgwidth{\linewidth}
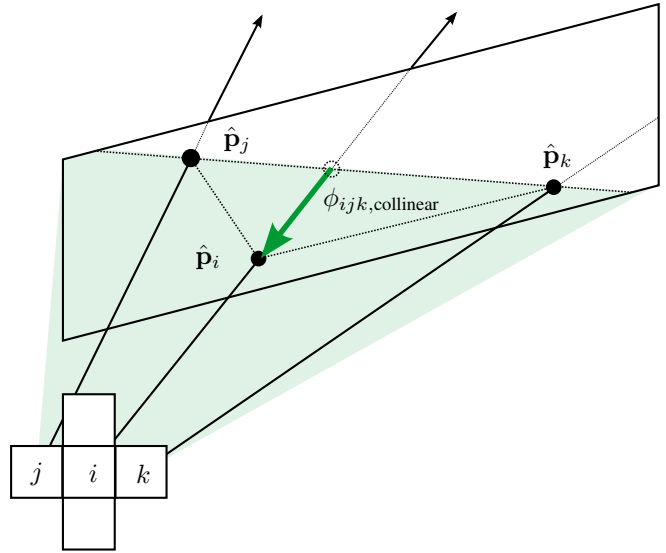
\caption{Collinearity residual computation.
We assume the viewing rays corresponding to the neighbors $i$, $i$ and $k$ to be coplanar.
The intersection between this plane and an ideal plane forms a line.
We then evaluate whether the points $\hat{\mathbf{p}}_i$, $\hat{\mathbf{p}}_j$ and $\hat{\mathbf{p}}_k$ are collinear.
This residual can be evaluated horizontally and vertically.}
\label{fig:collinearity_residual}
\end{figure}

We are, however, not interested in computing normals directly, but instead finding a residual term that evaluates the planarity of surface points.
Here, we assume neighboring viewing rays along one row or column to be coplanar.
Although, this assumption might not hold for arbitrary camera models, it can be justified for a sufficiently small neighborhood around a reference image coordinate.
As the intersection between the plane spanned by these viewing rays and an ideal surface plane forms a line, we can add the residual
\begin{equation} \label{eq:collinearity_residual}
\mathbf{\phi}_{i j k, \text{collinear}} = \frac{\mathbf{\Delta}_{ji}}{\norm{\mathbf{\Delta}_{ji}}} - \frac{\mathbf{\Delta}_{ik}}{\norm{\mathbf{\Delta}_{ik}}}
\end{equation}
that evaluates whether triples of points are collinear.
As depicted in figure \ref{fig:collinearity_residual},
\begin{equation} \label{eq:point_difference}
\mathbf{\Delta}_{ji} = \hat{\mathbf{p}}_i - \hat{\mathbf{p}}_j = \mathbf{ray}_i(\hat{d}_i) - \mathbf{ray}_j(\hat{d}_j).
\end{equation}
and $\mathbf{\Delta}_{ik}$ are the pairwise differences between the points $i, j, k$, where $j, k \in \mathcal{D}$ and $i$ are coplanar.

Using the collinearity residual in equation (\ref{eq:collinearity_residual}), we can add one residual term for direct neighbors with the same image row and one term for neighbors with the same column.
As only three parameters are coupled within each residual, the problem sparsity is increased which leads to better convergence properties compared to explicitly estimating surface normals.

\subsection{Regularization weights} \label{subsec:weights}

The collinearity residual in equation (\ref{eq:collinearity_residual}) should only be applied to areas satisfying the assumption of planar surfaces.
To accomplish this, we use additional image features expressed as weights $w_{i}(\mathcal{D})$.

Here, we employ weights $w_{ij}$ that are defined between neighboring image coordinates $i$ and $j$.
For the collinearity residual, $\mathcal{D}$ consists of three image coordinates $\{i, j, k \}$ and we determine pairwise weights
\begin{equation}
w_{ij} = g(\Delta) = g(\mathbf{f}_i - \mathbf{f}_{j}).
\end{equation}
as components of the regularization weights
\begin{equation} \label{eq:weights}
w_i(\mathcal{D}) = w_{ij}w_{ik}
\end{equation}
added for each image coordinate $i$.

Pairwise weights $w_{ij}$ are composed of a scalar weighting function $g$ and image features $\mathbf{f}_i$ and $\mathbf{f}_j$.
The weighting function might be exponential, sigmoid, step or even constant, which means that local image features have no influence on the regularization cost $\Phi_{\text{reg}}$.
However, it is important to note that arbitrary features might be used as long as they provide information about scene planarity.

\subsection{Prior Estimation} \label{subsec:prior_estimation}

In our contribution, we do not focus on solving the optimization problem efficiently by analyzing the underlying problem structure.
Please refer to \cite{chen2014fast} or \cite{Schneider2016} for hints on implementation details.
Instead, we suggest an initialization method based on linear interpolation that significantly reduces optimization time and the number of iterations, respectively.

For our initialization method, projected 3D points need to be found for every query image coordinate.
Therefore, we generate a kd-tree as described in \cite{muja14} that includes the set of point projections within the image coordinate frame.
This kd-tree search structure quickly provides references to the nearest laser point projections for any query image coordinate.

\begin{figure}[htbp] 
\centering \def\svgwidth{\linewidth}
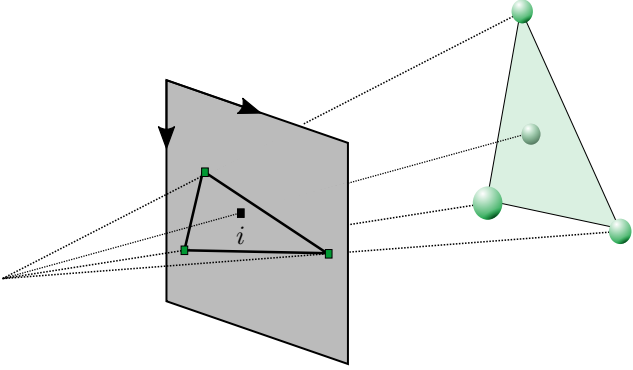
\caption{Initialization method based on intersecting the viewing ray of image coordinate $i$ with the plane constructed by a triangle mesh of the input point cloud.}
\label{fig:triangle}
\end{figure}

As depicted in Figure \ref{fig:triangle}, we generate a triangle mesh of the point cloud and project it into the image.
For each image coordinate $i$ within a triangle, we intersect the corresponding $\mathbf{ray}_i$ with the plane constructed by the three points defining the triangle which we use as depth initialization.
Image coordinates that are not covered by any triangle, will be assigned the depth of its nearest neighbor.
This might be the case at image borders or 3D points not connected by the meshing algorithm.

\subsection{Covariance Estimation}

In some scenarios, depth estimation may not work well.
On the one hand the range sensor's field of view might not cover the camera's field of view.
On the other hand, by evaluating equation (\ref{eq:weights}) image areas might be decoupled from their neighborhood and no depth observations exist in this area.
This might be the case when the image features of a pixel neighborhood indicate non-planar surfaces in a closed area and thus scale down regularization costs.

To resolve these problems, we aim to assign a confidence measure to each estimated depth after optimization by evaluating the covariance
\begin{equation}
\mathbf{C} = \left(\mathbf{J}'(\mathbf{d}^*) \cdot \mathbf{J}(\mathbf{d}^*)\right)^{-1},
\end{equation}
where the variances $\sigma^*_i$ can then be obtained by evaluating $\mathbf{C}(i, i)$. 

Knowing an estimate $\sigma^*_i$ for $d^*_i$ we can then set a threshold and keep only those distances with variances below that threshold.

\subsection{Implementation}

Equations (\ref{eq:data_cost}) and (\ref{eq:regularization_cost}) show that the Markov Random Field formulation can be expressed as a nonlinear least-squares problem for which we aim to find to optimal parameters, i.e. parameters that minimize the overall costs.
The problem consists of many residual terms, each of them depending on either one or three parameters.
In total, we add one residual term for each depth observation and approximately two residual terms for each pixel if image borders are disregarded.
The resulting problem can then be solved by Trust-Region methods using a linear solver efficiently exploiting the sparse problem structure.

We implemented our Markov Random Field-based upsampling method as a C++ library which will be publicly available on \url{https://github.com/fzi-forschungszentrum-informatik/mrf}.
It is based on Ceres Solver \cite{ceres}, an optimization framework used to solve large-scale, non-linear least squares problems.
As residual blocks can be added one by one, Ceres itself exploits the sparse structure and uses state-of-the-art sparse linear solver libraries in its backend.
Additionally, parameters can be constrained on minimum or maximum bounds that we set to the minimum and maximum depth observed by the range sensor.

\section{Applications and Experiments}
\label{sec:applications_and_experiments}

In section \ref{subsec:scenenet_rgb-d}, we introduce our performance metrics and show evaluation results on a photorealistic RGB-D dataset.

We then present our experimental platform for the mapping of urban environments as an application of guided upsampling and perform a qualitative evaluation in section \ref{subsec:experimental_vehicle}.

For both applications, we compare our approach to the model presented in \cite{Thrun2005} where a constant distance regularization is used.

\subsection{Photorealistic Indoor Dataset} \label{subsec:scenenet_rgb-d}

We evaluate our approach on a subset of 150 images of the \textit{SceneNet RGB-D} \cite{scenenet} dataset.
It provides RGB-D sensor data from photo-realistic synthetic indoor scenes which are semantically labelled by instances and a camera model.

\begin{figure}[b]
\centering
\def\svgwidth{\linewidth}
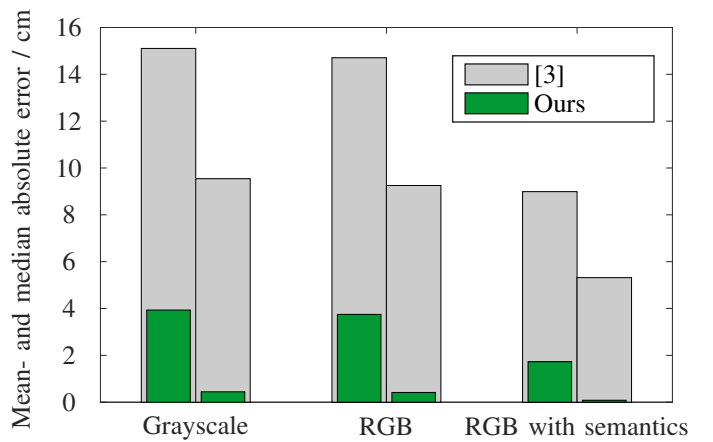
\caption{Mean and median absolute error depending on different image features used.
Semantic features drastically improve the upsampling quality.}
\label{fig:comparison_features} 
\end{figure}

In the dataset ground-truth depth information $d_i$ is available for each estimated depth $\hat{d}_i$ for all image coordinates $i$ in the image $\mathcal{I}$.
Here, we determine the mean
\begin{equation}
\frac{1}{\vert\mathcal{I}\vert} \sum_{i \in \mathcal{I}} \vert e_i \vert
\end{equation}
and the median of the absolute error $\vert e_i \vert = \vert \hat{d}_i - d_i \vert$.
For each evaluation, we also provide the downsampling ratio
\begin{equation}
r = \frac{\vert \mathcal{O} \vert}{\vert \mathcal{I} \vert}
\end{equation}
which is defined by the amount of 3D observations divided by the image size.

Figure \ref{fig:comparison_features} shows the upsampling quality for different image features used.
We observe that semantic features drastically improve the upsampling quality.
Our method achieves an mean absolute error of about 17 mm and an even lower median absolute error if RGB and semantic features are used.
Here, the average downsampling rate is 1$\%$.

\begin{figure}[t]
\centering
\def\svgwidth{\linewidth}
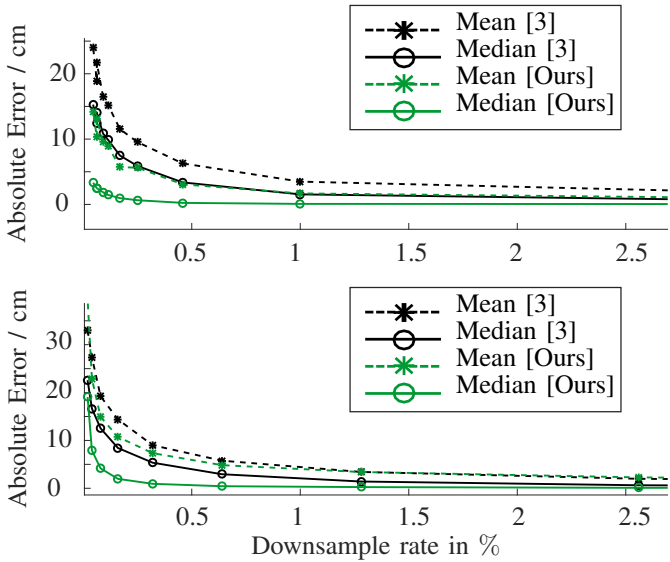
\caption{Mean and median absolute error depending on the downsampling ratio.
Top: Equidistant downsampling, bottom: random downsampling}
\label{fig:downsampling_ratio} 
\end{figure}

Figure \ref{fig:downsampling_ratio} depicts the absolute errors depending on different downsampling ratios, i.e. the sparsity of 3D observations.
For our evaluations, we performed equidistant as well as random downsampling.
Whereas the mean absolute errors are comparable for a larger number of observations, our approach outperforms for few observations.
The reason might be a more realistic regularization in scenes containing a moderate amount of planar surfaces.

\subsection{Experimental Vehicle} \label{subsec:experimental_vehicle}

\begin{figure}[htbp]
\centering \def\svgwidth{\linewidth}
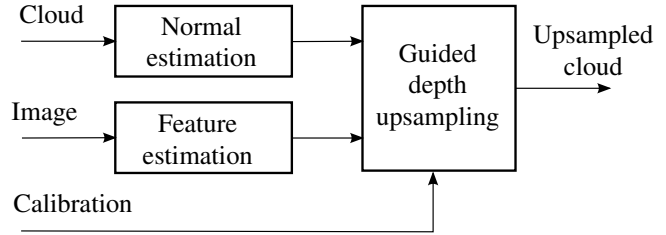
\caption{System overview.
Before upsampling, surface normal features for laser data and semantic image features are estimated.
The upsampled cloud contains surface normals, image features and the depth confidence measure.}
\label{fig:system_overview}
\end{figure}

Figure \ref{fig:system_overview} depicts the upsampling pipeline implemented for our experimental vehicle.

Our platform is equipped with a Velodyne HDL64E-S2 lidar and a high definition RGB camera.
The lidar is mounted on top of the vehicle to generate range sensor data structured as a 3D point cloud.
RGB images are provided by a Teledyne Dalsa Genie TS-C4096 color camera with an approximate resolution of 12 Megapixels which is mounted externally above the windshield.
Camera and laser are triggered at the same rate and the pose between laser scanner and camera can be assumed calibrated to an accuracy of $\pm 0.5\deg$.
For one scenario, the projection of laser points into the camera image is depicted in the top image of Figure \ref{fig:upsampled_real_scene}.

\begin{figure*}[!t]
\centering
\includegraphics[width=\textwidth]{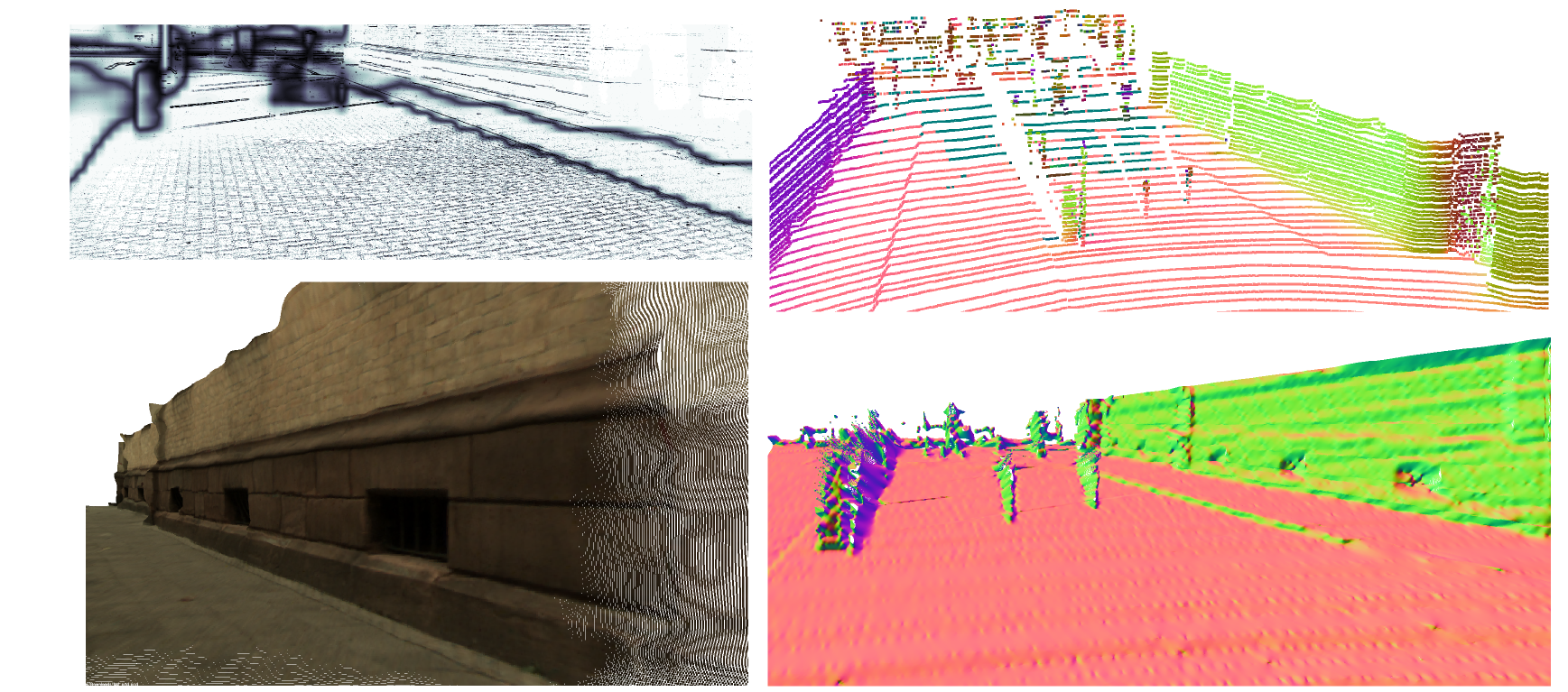}
\caption{\textit{Top left}: Regularization weights determined from RGB values and semantic output certainty,
\textit{top right}: Based on the sparse input cloud normals are estimated,
\textit{bottom left}: Building front in the upsampled scene,
\textit{bottom right}: Normals within the upsampled cloud are mostly smooth}
\label{fig:collection}
\end{figure*}

Based on 3D point cloud information provided, we estimate surface normals similar to \cite{pcl_normal} for each point observed.
The method is based on a Principle Component Analysis of all points within a search radius around a query point.
The search radius might be adapted depending on the range sensor model.
We assign the Eigenvector corresponding to the smallest Eigenvalue to the surface normal of that query point.
These surface normals may then be included as pseudo measurements into our guided depth upsampling system.
An exemplary normal estimation result is depicted on the top right corner of Figure \ref{fig:collection}.

Our image features
\begin{equation}
\mathbf{f}_i = \begin{bmatrix}\mathbf{f}_{i, \text{rgb}} \\ f_{i, \text{semantic}}\end{bmatrix}
\end{equation}
are composed of the RGB value $\mathbf{f}_{i, \text{rgb}}$ and a semantic certainty $f_{i, \text{semantic}}$.
Therefore, we predict semantic classes using GoogLeNet \cite{Szegedy_2015_CVPR} adapted as FCN-8s \cite{long2015fully}.
The network was trained on a 14-class subset of the cityscapes dataset \cite{cordts2016cityscapes}.
Apart from the arg-max class predictions, we utilize the semantic certainty
\begin{equation}
f_{i, \text{semantic}} = (Np_i - 1)/(N-1) \in [0, 1],
\end{equation}
of that class, where $N$ is the number of classes.
It is computed from the network's softmax output $p_i$, i.e. the output's improvement over guessing normalized to the maximum possible improvement.
For certain predictions this value becomes 1 while at class boundaries, it drops to 0. 

Finally, pair-wise weights
\begin{equation}
w_{ij} = g(\mathbf{f}_i, \mathbf{f}_j) = g(\norm{\mathbf{f}_{i, \text{RGB}} - \mathbf{f}_{j, \text{RGB}}}^2 \cdot f_{i, \text{semantic}}
\end{equation}
are calculated where a scaling function $g$ is applied to the difference in the RGB space between pixel $i$ and $j$, weighted with the semantic class certainty at pixel $i$.
The regularization weights $w_{ijk}$ as applied in equation (\ref{eq:weights}) are depicted in the top left corner of Figure \ref{fig:collection}.

Based on regularization weights, camera model and 3D surface point normals, upsampling is performed.
The upsampled depth image for this scenario is depicted on the bottom of figure \ref{fig:upsampled_real_scene}.
Using the ray lookup function in equation (\ref{eq:point_from_depth}), we can transform this depth image into a 3D point cloud which is depicted for a shifted viewpoint on the bottom of Figure \ref{fig:collection}.
We observe that for a soft regularization scaling due to the semantic certainty, some objects in the scene are not completely separated from the environment.
However, our approach accurately estimates planar surfaces such as house fronts or ground surfaces.

\section{Conclusion}
\label{sec:conclusion}

We presented an approach for guided depth upsampling of range sensor data based on a novel regularization term that preserves plane surfaces.
Furthermore, we do not only incorporate 2D image features into our model but also 3D surface normals.
By using a novel regularization term evaluating surface planarities, we show that our method outperforms state-of-the-art methods regularizing towards constant depths.
Finally, we suggest a method to filter ill-conditioned data based on estimating the covariance matrix after optimization.
As the upsampling quality is sensitive to calibration and synchronization errors, we would also like to include the transformation between laser and camera into the optimization problem which might lead to a one-shot extrinsic calibration technique.


\bibliographystyle{IEEEtran/IEEEtran}
\bibliography{refs}

\end{document}